\documentclass[aps,english,twocolumn,prb]{revtex4}
\usepackage[T1]{fontenc}
\usepackage[latin9]{inputenc}
\usepackage{amsmath}
\usepackage{graphicx}
\makeatletter

\@ifundefined{textcolor}{}
{%
 \definecolor{BLACK}{gray}{0}
 \definecolor{WHITE}{gray}{1}
 \definecolor{RED}{rgb}{1,0,0}
 \definecolor{GREEN}{rgb}{0,1,0}
 \definecolor{BLUE}{rgb}{0,0,1}
 \definecolor{CYAN}{cmyk}{1,0,0,0}
 \definecolor{MAGENTA}{cmyk}{0,1,0,0}
 \definecolor{YELLOW}{cmyk}{0,0,1,0}
 }

\makeatother

\usepackage{babel}
\begin{document}

\title{Ordinary Percolation with Discontinuous Transitions}
\author{Stefan Boettcher and Vijay Singh}
\affiliation{Department of Physics, Emory University, Atlanta, GA 30322; USA}
\author{Robert M. Ziff}
\affiliation{Center for the Study of Complex Systems and Department of Chemical
Engineering, University of Michigan, Ann Arbor, MI 48109-2136; USA}
\begin{abstract}
Percolation on a one-dimensional lattice and fractals such as the
Sierpinski gasket is typically considered to be trivial because they
percolate only at full bond density. By dressing up such
lattices with small-world bonds, a novel percolation transition with
explosive cluster growth can emerge at a nontrivial critical point.
There, the usual order parameter, describing
the probability of any node to be part of the largest cluster, jumps
instantly to a finite value. Here, we provide a simple example of this
transition in form of a small-world network consisting of a
one-dimensional lattice combined with a hierarchy of long-range bonds
that reveals many features of the transition in a mathematically
rigorous manner.
\end{abstract}
\maketitle

\section*{Introduction}
The percolation properties\citep{Stauffer94} of networks are of
significant interest --- without percolation, any large-scale
transport or communication through the network ceases. Much research
has been dedicated to the understanding of percolation on randomly
grown, complex networks\citep{Barabasi03}.  Yet, the engineering of
artificial networks with well-controlled features seems
desirable. Indeed, there has been considerable interest in the
properties of spatial networks, linking real-world geometry with
small-world
effects\citep{Watts98,Boguna09,barthelemy_spatial_2010}. In
particular, networks possessing hierarchical
features\citep{Trusina04,Andrade05,Hinczewski06,SWPRL,Boguna09,PhysRevE.82.036106}
relate to actual transport systems such as for air travel, routers,
and social interactions.  Certain hierarchical networks with a
self-similar structure have been shown to exhibit novel
features\citep{Hinczewski06,Boettcher09c,Berker09,Nogawa09,Hasegawa10a,Boettcher10c}.
To these, we add here an unprecedented discontinuous transition in the
formation of an extensive cluster for ordinary, random bond
addition. Such an extensive cluster is said to percolate, as it
contains a finite fraction of all nodes. Remarkably, this
discontinuous transition can be derived \emph{exactly} with recursive
methods, as shown below.

For a random network that allows bonds between any pair of nodes, the
possibility of a discontinuous (``explosive'') percolation transition
has recently attracted considerable
attention\citep{Achlioptas09,ISI:000286751500010,ISI:000291093600009,ISI:000287844300028,PhysRevE.84.020101,ISI:000279888400004,FriedmanLandsberg09,RozenfeldGallosMakse09,RadicchiFortunato10,ChoKahngKim10,ISI:000277699600032,ISI:000288007800002,MannaChatterjee11}. Such
a transition raises the prospect that a minute increase in the
bond-density $p$ of a network can suddenly make a large fraction of
all nodes accessible, for instance, for the spreading of a
contagion\citep{Balcan09}.  Yet, all proposed mechanisms for such a
transition require \emph{correlated} bond
additions\citep{Achlioptas09}. While further simulations of the
dynamics of such correlated cluster
formation\citep{ISI:000279888400004,FriedmanLandsberg09,RozenfeldGallosMakse09,RadicchiFortunato10,ChoKahngKim10,ISI:000277699600032,ISI:000288007800002,MannaChatterjee11}
seemed to support the existence of a discontinuity, evidence against
it\citep{ISI:000286751500010,ISI:000291093600009,ISI:000287844300028,PhysRevE.84.020101}
finally mounted into a general proof of the impossibility of explosive
percolation for any of the proposed mechanisms\cite{Riordan11}. In
contrast to these efforts, we study ordinary (uncorrelated) bond
additions but on networks with a recursive, hierarchical structure to
induce such an explosive percolation dynamics.  Understanding of
explosive cluster formation will be significantly advanced when the
discontinuity can be studied rigorously by simply adding bonds
randomly to these networks.

\begin{figure}
\includegraphics[clip,scale=0.30]{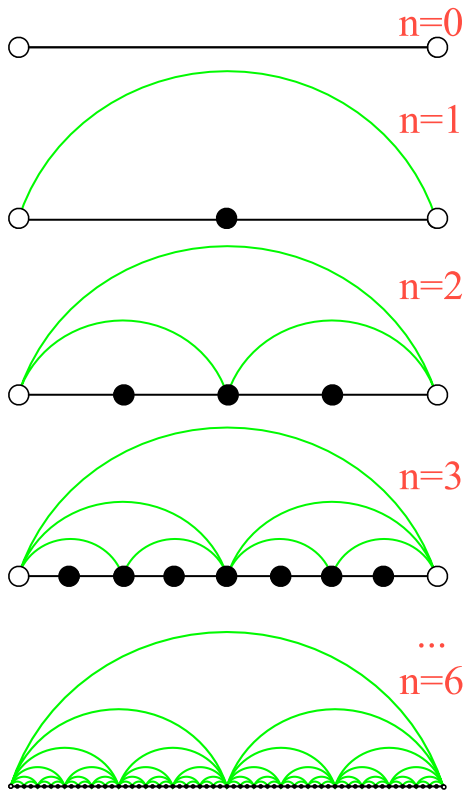}
\caption{\label{fig:MK1}{\bf Recursive  generation of a
  hierarchical network.} (The network is displayed at full bond density,
  \emph{$p=1$}.) In $(n=0)$-th generation, the network consists of a
  single bond between two end-nodes (open circles). The probability
  for an end-to-end path is obviously $T_{n=0}=p$. In successive
  generations $n+1=1,2,\ldots$, two sub-networks from the prior
  generation $n$ are merged together with the endpoints connected by a
  new long-range bond (shaded arcs).  There is an end-to-end path with
  probability $T_{n+1}$, if either the new long-range bond exists
  (probability $p$, irrespective of $T_{n}$) or both prior
  sub-networks are present (probability $(1-p)\, T_{n}^{2}$), leading
  to Eq.~(\ref{eq:RGrecurMK1}).  In each generation $n$ the network
  has $N=2^n+1$ nodes and $E=2^{n+1}-1$ bonds for an average degree
  $2E/N\sim4$ for large $n$.  }
\end{figure}

\section*{Results}
\paragraph*{\bf Probability of a Spanning Cluster.}
In our discussion we focus on a simple example of a hierarchical
network that is developed in Fig.~\ref{fig:MK1}. In its hierarchical
construction, networks from preceding generations are merged for
successive generations for ever larger networks.  The probability for
an end-to-end path of the most elementary network at $n=0$, a single
bond, is $T_{n=0}=p$. By merging two networks of generation $n$
side-by-side and adding a long-range bond to obtain a network of
generation $n+1$, we recursively determine the probability
$T_{n+1}$. This analysis is but one example of the real-space
renormalization group (RG)\citep{Plischke94}.

The probabilities for an end-to-end path satisfy the recursion
\begin{eqnarray}
T_{n+1} & = & p+(1-p)T_{n}^{2},\qquad\left(T_{0}=p\right),
\label{eq:RGrecurMK1}
\end{eqnarray}
as is explained in Fig.~\ref{fig:MK1}.  In the limit of infinitely
large networks we obtain for any $p$ the probability for such an
end-to-end connection, $T^*(p)=\lim_{n\to\infty}T_n$, from the
stationary solutions of Eq.~(\ref{eq:RGrecurMK1}), called \emph{fixed
  points}. Setting $T_{n+1}\sim T_n\sim T^*$ in
Eq.~(\ref{eq:RGrecurMK1}) yields a horizontal line of fixed points
$T^{*}\equiv1$ for all $p$, as well as arising line of fixed points,
\begin{eqnarray}
T^{*}(p) & = & \frac{p}{1-p}\qquad\left(p\leq\frac{1}{2}\right).
\label{eq:RG_MK1_fp}
\end{eqnarray}
Both lines intersect at $p=\frac{1}{2}$. Further analysis
shows\citep{Boettcher09c} that Eq.~(\ref{eq:RG_MK1_fp}) is the stable
solution for all $0<p<\frac{1}{2}$ that describes the actual behavior
of large networks. I.e., very large networks possess an end-to-end
connection with a finite probability approaching
$0<T^{*}(p)<1$. However, for $\frac{1}{2}\leq p\leq1$, the horizontal
line is the only physical and stable solution such that a connection
exists with certainty, $T^{*}=1$.

\paragraph*{\bf Origin of the Percolation Transition.}
The phenomenology of percolation on hierarchical networks is quite
distinct from that of
lattices\citep{Boettcher09c,Nogawa09,Berker09,Hasegawa10a}.
Specifically, on a lattice both, an end-to-end path and an extensive
cluster, arise with certainty above the same critical bond density
$p_c$. Each node attains a finite probability
$P_{\infty}\left(p>p_c\right)>0$ to be connected to the spanning
cluster. In contrast, hierarchical networks may have \emph{two}
transitions at a lower and an upper bond density, $p_{l}<p_{u}$. Below
the lower transition $p_{l}$, all clusters remain finite and no
end-to-end paths exists. Above the upper transition $p_{u}$, there is
an extensive cluster and a certain end-to-end path, as on any
lattice. But both transitions delimit an interval
$p\in\left(p_{l},p_{u}\right)$ that contains fractal
(sub-extensive) clusters with a finite probability $T^*$ for an
end-to-end path, as given by Eq. (\ref{eq:RG_MK1_fp}). These clusters
each harbor a vanishing fraction of all nodes, however, they diverge
in size with $N^\Psi$, defining a new fractal exponent
$0\leq\Psi\left(p\right)<1$\citep{Nogawa09}.  Accordingly, the order
parameter\citep{Stauffer94,Plischke94} $P_{\infty}$ becomes non-zero
only at $p_{u}$, making it the thermodynamically correct transition
point, $p_{c}=p_{u}$.  In the present network it is $p_{l}=0$ and
$p_u=\frac{1}{2}$; more elaborate networks with exact but non-trivial
$p_{l}$ and $p_{u}$ are discussed elsewhere\citep{Boettcher09c}.

We note that the simultaneous emergence of an extensive cluster with
end-to-end paths appears to be special for ``flat''
geometries. Hierarchical networks posses what is called a hyperbolic
geometry\citep{PhysRevE.82.036106} for which most nodes are close to
the periphery, similar to a tree, with many root-to-end
paths. Correspondingly, parabolic networks would be very prone to
clustering in the bulk with few paths to any peripheral node, i.e., an
extensive cluster would emerge before paths that access the periphery
arise.

\begin{figure}
\begin{centering}
\includegraphics[bb=50bp 350bp 700bp 550bp,clip,scale=0.32]{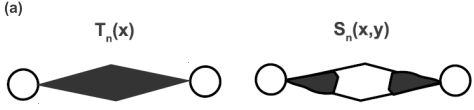} 
\par\end{centering}

\caption{\label{figMK1GenFunc}{\bf Diagramatic defintion of the
    generating functions.} In the schematic for generating functions
  $T_{n}(x)$ and $S_{n}(x,y)$, the open circles represent
  the end-nodes, shaded areas indicate clusters that either span
  ($T_{n}$) or do not span ($S_{n}$) between the end-nodes. Clusters
  that do not reach an end-node are ignored. }
\end{figure}

\paragraph*{\bf Construction of the Order Parameter.}
To reveal the nature of the transition, the probability $T^*(p)$ alone
provides insufficient information. In addition, we have to derive the
average size $\left\langle s_{\rm max}\right\rangle_n$ of the largest
cluster and
\begin{eqnarray}
P_{\infty}(p)=\lim_{n\to\infty}\frac{\left\langle s_{{\rm  max}}\right\rangle _n}{N}
\label{Pinfty}
\end{eqnarray}  
as the proper order parameter\citep{Stauffer94} from cluster
generating functions. Thus, we introduce two basic quantities: the
probability $t_{i}^{(n)}(p)$ that both endnodes are connected to the
same cluster of size $i$, and the probability $s_{i,j}^{(n)}(p)$ that
the left endnode is connected to a cluster of size $i$ and the right
endnote to a different cluster of size $j$.  The corresponding
generating functions to provide the average cluster size are defined
as
\begin{eqnarray}
T_{n}(x) & = &\sum_{i=0}^{\infty}t_{i}^{(n)}(p)x^{i},\label{T_n}\\ 
S_{n}(x,y) & = &\sum_{i=0}^{\infty}\sum_{j=0}^{\infty}s_{i,j}^{(n)}(p)x^{i}y^{j}
\label{S_n}
\end{eqnarray}
and are depicted in Fig~\ref{figMK1GenFunc}. $T_n(x)$ represents
clusters with an effective bond between the endnodes while $S_n(x,y)$
represents those that fail to provide such a bond. The following
manipulations, while subtle, only entail elementary manipulations that
result in three coupled but linear recursions. Although we provide
enough details here to reproduce the intermediate steps in a few
lines, we have implemented those steps also as a Mathematica script
provided in the Supplementary Software. Therein lies the advantage of
our present example: The discontinuity can be obtained exactly and
with ease.

\begin{figure}
\begin{centering}
\includegraphics[clip,scale=0.3]{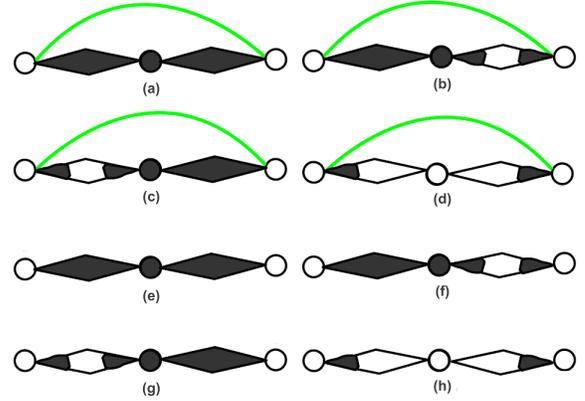}
\par\end{centering}

\caption{\label{figMK1GenFuncb}{\bf Diagramatic evaluation of the
    generating functions.} 
  All diagrams contributing
  to $T_{n+1}(x)$ or $S_{n+1}(x,y)$ in the $n$th RG step are shown.
  The remaining end-nodes (always-open circles) are not
  counted in the generating functions; the (black) connecting nodes
  increment the cluster size, accounted for by a factor of $x$ or
  $y$. The contribution of each configuration is: (a)
  $xpT_{n}^{2}(x)$, (b) $xpT_{n}(x)S_{n}(x,x)$, (c)
  $xpT_{n}(x)S_{n}(x,x)$, (d) $pS_{n}(x,1)S_{n}(1,x)$, (e)
  $x(1-p)T_{n}^{2}(x)$, (f) $x(1-p)T_{n}(x)S_{n}(x,y)$, (g)
  $y(1-p)T_{n}(y)S_{n}(x,y)$, and (h) $(1-p)S_{n}(x,1)S_{n}(1,y)$.
  As tallied up in Eqs.~(\ref{eq:T_n+1_old},\ref{eq:S_n+1_old}),
  configurations (a)-(e) span end-to-end and contribute to
  $T_{n+1}(x)$, while (f)-(h) do not span and contribute to
  $S_{n+1}(x,y)$.}
\end{figure}

The recursion relations for these generating functions are obtained in
Fig.~\ref{figMK1GenFuncb} by considering all possible configurations
on three nodes, similar to Fig.~\ref{fig:MK1} but now also taking
cluster sizes into account.  For each configuration, the type of
effective bond between each of the nodes is checked, see
Fig.~\ref{figMK1GenFunc}, and these bonds are assigned a value
$T_{n}(x)$ or $S_{n}(x,y)$, depending on the type of cluster each
represents. As in Fig.~\ref{fig:MK1}, small world bonds are merely
assigned the probability $p$ or $1-p$ for being present or
not. Marking the increment in cluster size, the inner node provides a
factor of $x$ or $y$ for its adjacent cluster, or unity if it remains
isolated. The contribution of each configuration to the next generation is
the product of the weights of the three bonds and of the intermediate
node; all eight of these are determined in Fig.~\ref{figMK1GenFuncb}.
From Fig.~\ref{figMK1GenFuncb} we read off the recursions
\begin{eqnarray}
T_{n+1}(x) & = & xT_{n}^{2}(x)+p\left[2xT_{n}(x)S_{n}(x,x)\right.\nonumber \\
 & {} & \left.+S_{n}(x,1)S_{n}(1,x)\right],\label{eq:T_n+1_old}\\
S_{n+1}(x,y) & = & (1-p)\left[xT_{n}(x)S_{n}(x,y)+\right.\nonumber \\
 & {} & \left.+yT_{n}(y)S_{n}(x,y)+S_{n}(x,1)S_{n}(1,y)\right],\label{eq:S_n+1_old}
\end{eqnarray}
initiated with $T_{0}(x)=p$, $S_{0}(x,y)=1-p$. Naturally, both
equations reduce to Eq.~(\ref{eq:RGrecurMK1}) for $x=y=1$ where
$T_{n}\left(1\right)=1-S_{n}\left(1,1\right)=T_{n}$.

The recursions in Eqs.~(\ref{eq:T_n+1_old},\ref{eq:S_n+1_old}) contain
more information then is needed here and we simplify them in terms of
functions of a single variable $x$. We define the functions
$\Sigma_{n}(x)\equiv S_{n}\left(x,x\right)$ and ${\cal S}_{n}(x)\equiv
S_{n}\left(x,1\right)=S_{n}\left(1,x\right)$ that, combined into a
more efficient vector notation $\vec{V}=\left[T, \Sigma, {\cal
    S}\right]$, lead to
\begin{equation}
\vec{V}_{n+1}\left(x\right)=\vec{{\cal F}}\left(\vec{V}_{n}\left(x\right),x\right)
\label{eq:V_n+1}
\end{equation}
with a function-vector $\vec{{\cal F}}$ of non-linear components
\begin{eqnarray}
\vec{{\cal F}}\left(\vec{V},x\right) & = & \left[\begin{array}{r}
xT^{2}+2xpT\Sigma+p{\cal S}^{2}\\
(1-p)\left(2xT\Sigma+{\cal S}^{2}\right)\\
(1-p)\left(1+xT\right){\cal S}
\end{array}\right].\label{eq:vecF}
\end{eqnarray}

As needed in Eq.~(\ref{Pinfty}), the mean size $\left\langle s_{{\rm
    max}}\right\rangle _n$ of the cluster connected to the endnodes
results from the first moment of the generating functions. These are
obtained via their first derivative in $x$ at $x=1$,
\begin{equation}
\left\langle s_{{\rm max}}\right\rangle_n=
T_n^{\prime}\sim N^{\Psi\left(p\right)},
\label{eq:s_max}
\end{equation}
for a network of size $N=2^n+1\to\infty$. 

\paragraph*{\bf Analysis of the Recursions for the Mean Cluster Size.}
The mean size of the largest cluster, as needed in Eq.~(\ref{Pinfty}) to
construct the order parameter, is obtained
by Taylor-expanding Eq.~(\ref{eq:V_n+1}) to first order in
$\epsilon\equiv1-x\to0$.  To zeroth order, each component in
Eq.~(\ref{eq:V_n+1}) evaluated at $x=1$ reproduces
Eq. (\ref{eq:RGrecurMK1}) again. To order $\epsilon$, we find an
linear inhomogeneous recursion for
$\vec{V}_{n}^{\prime}$, dropping the now-redundant argument $x=1$,
\begin{eqnarray}
\vec{V}_{n+1}^{\prime} & = & \frac{\partial\vec{{\cal F}}}{\partial\vec{V}}\left(\vec{V}_{n}\right)\circ\vec{V}_{n}^{\prime}+\frac{\partial\vec{{\cal F}}}{\partial x}\left(\vec{V}_{n}\right)\label{eq:Vprime}
\end{eqnarray}
with the Jacobian matrix 
\begin{equation}
\frac{\partial\vec{{\cal F}}}{\partial\vec{V}}\left(\vec{V}\right)=
\begin{bmatrix}
2T+2p\Sigma,           & 2pT,             & 2p{\cal S}\\
2(1-p)\Sigma,           & 2(1-p)T,       & 2(1-p){\cal S}\\
(1-p){\cal S},             & 0,                  & (1-p)\left(1+T\right)
\end{bmatrix}
\label{eq:3matrix}
\end{equation}
and the inhomogeneity from differentiating for $x$ explicitly
\begin{eqnarray}
\frac{\partial\vec{{\cal F}}}{\partial x}\left(\vec{V}\right) & = & \left[\begin{array}{r}
T^{2}+2pT\Sigma\\
2(1-p)T\Sigma\\
(1-p)T{\cal S}
\end{array}\right].
\label{eq:inhom}
\end{eqnarray}
In Eqs.~(\ref{eq:3matrix}) and~(\ref{eq:inhom}) we neglected the index
$n$ on $\vec{V}$ and its components to simplify the presentation. They
depend on $n$ through $\vec{V}_{n}=\left[T_{n},\Sigma_n=1-T_{n},{\cal
    S}_n=1-T_{n}\right]$.  Since each network at $n=0$ only consists
of endnodes, which are not counted, all clusters are initially empty, i.e.,
$\vec{V}_{0}^{\prime}=\left[0,0,0\right]$.

For large $n$ at $x=1$, i.e., near the fixed point
$\vec{V}^{*}=\left[T^{*},\Sigma^*=1-T^{*},{\cal S}^*=1-T^{*}\right]$,
it is easy to show that the inhomogeneity in Eq.~(\ref{eq:Vprime}) is
subdominant, leaving a linear homogeneous system with constant
coefficient-matrix $\frac{\partial\vec{{\cal
      F}}}{\partial\vec{V}}\left(\vec{V}^{*}\right)$.  The largest
eigenvalue $\lambda$ of this matrix provides the dominant contribution
for each component of $\vec{V}_{n}^{\prime}$, i.e.,
$T_n^{\prime},\Sigma_n^{\prime},{\cal S}_n^{\prime}\sim\lambda^n$.  We
obtain $\lambda$ between the transitions, $0=p_l\leq
p<p_u=\frac{1}{2}$, by applying Eq.~(\ref{eq:RG_MK1_fp}) for $T^{*}$
in the matrix. Via Eq.~(\ref{eq:s_max}) it is $\langle s_{{\rm
    max}}\rangle_n\sim\lambda^n$ for $n\to\infty$, which yields
the fractal exponent
\begin{equation}
\Psi(p)=\frac{\ln\lambda}{\ln2},\quad  
\lambda=\frac{1+3p-4p^2}{2(1-p)}+\sqrt{\frac{1-p(1-4p)^2}{4(1- p)}}.
\label{eq:PsiValue}
\end{equation}
The largest eigenvalue always remains $\lambda<2$  for $p<1/2$, i.e.,
$0\leq\Psi(p)<1$, which implies that the order parameter $P_{\infty}$ in
Eq.~(\ref{Pinfty}) vanishes for $p<p_u$, hence, $p_u=p_c$.

\begin{figure}
\includegraphics[clip,scale=0.3]{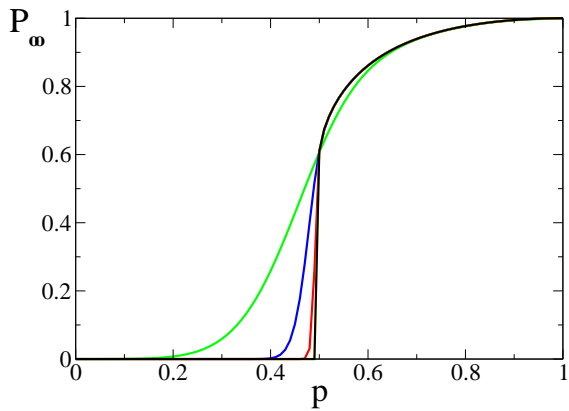}
\caption{\label{fig:PinftyMK1}{\bf Plot of the order parameter
    $P_{\infty}\left(p\right)$ in Eq.~(\ref{Pinfty}).}
  $P_{\infty}\left(p\right)$ is evaluated after $n=10^k$
  iterations of the recursions in Eq.~(\ref{eq:Vprime}) with
  $k=1,\ldots,5$, displayed from left to right. This
  corresponds to system sizes of up to $N\approx2^n\sim10^{3010}$
  nodes. It evolves slowly into a discontinuity at $p\to p_{c}=\frac{1}{2}$
  with $P_{\infty}\left(p_{c}\right)=0.609793\ldots$.  Convergence is
  slowest just below $p_{c}$, since finite-size corrections decay as
  $N^{\Psi\left(p\right)-1}$ with
  $1-\Psi\left(p\right)\sim8\left(p_{c}-p\right)^2/\ln2$ for $p\to
  p_{c}$ from Eq.~(\ref{eq:PsiValue}).}
\end{figure}

Above and \emph{at} the transition, $p_{c}\leq p\leq1$, it is
$T^{*}=1$ and Eq.~(\ref{eq:3matrix}) provides uniformly $\lambda=2$ as
the largest eigenvalue (i.e., $\Psi\equiv1$), indicating percolation
in form of an extensive cluster. For a continuous transition,
$P_{\infty}\left(p\right)\sim\left(p-p_c\right)^\beta\to0$ with
$\beta>0$ for $p\to p_{c}^{+}$. In contrast, Eq.~(\ref{eq:Vprime}) can
be shown rigorously to provide a monotone increasing sequence for
$T_n^{\prime}$, exactly at $p=p_{c}=\frac{1}{2}$ and for any $p$ above
(see Supplementary Software).  Therefore, the order parameter is
positive definite even exactly at $p=p_c$, as displayed in
Fig.~\ref{fig:PinftyMK1}. In fact, the continuity of
$P_{\infty}\left(p\right)$ is interrupted merely because
$T^{*}\equiv1$ suddenly becomes an \emph{unstable} fixed point of
Eq.~(\ref{eq:RGrecurMK1}) below $p_{c}$. There, the stable branch
transitions to Eq.~(\ref{eq:RG_MK1_fp}) that lacks extensive clusters,
$\Psi<1$. Hence, it is the intersection of two stable branches of
fixed points $T^{*}$ at $p_u$ that causes a discontinuous
transition. Such intersections of lines of fixed points is generic in
hierarchical networks\citep{Boettcher09c}, whereas fixed points for
percolation on lattices always remain \emph{isolated}.

\section*{Discussion}
We have shown that a hierarchy of small-world bonds grafted onto a
one-dimensional lattice can result in an explosive percolation
transition, even if bonds are added sequentially in an uncorrelated
manner. The discontinuous transitions found in hierarchical networks
are unique as alternative models based on correlated bond additions
have been proven to fail\cite{Riordan11}.  At this point, the precise
conditions to be imposed on the hierarchy of long-range bonds for
obtaining this transition are not entirely clear. However, in each
example we have obtained the addition of small-world bonds converted
an initially finitely-ramified network into an infinitely ramified
network to provide $p_{c}<1$, as several other networks
demonstrate\citep{Boettcher09c}. In a finitely-ramified network, by
definition\citep{Stauffer94}, the removal of just a finite number of
bonds can separate off extensive clusters in the limit of large
systems, $N\to\infty$, resulting in $p_c=1$.  In contrast, studies of
hierarchical systems with small-world bonds imposed on apriori
infinitely-ramified
2d-lattices\citep{Boettcher09c,Berker09,Hasegawa10a} appear to result
in infinite-order transitions instead, which have been observed in
many other networks\citep{Dorogovtsev08}.

\section*{Acknowledgements}
SB thanks M.~Paczuski, P.~Grassberger, and the entire Complex Science
Group at University of Calgary for helpful discussions. This work has
been partially supported by grant DMR-0812204 from the National
Science Foundation.

\section*{Author contributions}
S.B.~and R.M.Z.~developed the idea for this research, S.B.~and
V.S.~worked out the formalism, and V.S.~implemented the formalism,
interpreted the results, and conducted numerical tests in close
collaboration with S.B. 

\section*{Competing Financial Interest Statement}
None of the authors has any competing financial interests arising from any content in this paper.

\bibliographystyle{unsrt}
\bibliography{/Users/stb/Boettcher}

\end{document}